\documentclass[journal,10pt,twocolumn]{IEEEtran}
\usepackage{comment}
\usepackage{amsmath,amssymb,amsthm}

\usepackage{cite}
\usepackage{pdflscape}
\usepackage[english]{babel}
\usepackage{multicol}
\usepackage{multirow}
\usepackage{algorithm}
\usepackage[noend]{algpseudocode}

\usepackage{amsmath}
\usepackage{amsmath}
\usepackage{amsfonts}
\usepackage{amssymb}
\usepackage{float}
\theoremstyle{definition}
\newtheorem{defn}{Definition}[]
\usepackage{stfloats}
\usepackage{subfigure}
\usepackage{float}
\usepackage{amsmath}
\usepackage{mathrsfs}
\usepackage{color}
\usepackage{float}
\usepackage{array}
\usepackage{balance}
\usepackage{graphicx}
\usepackage{epsfig}
\usepackage{epstopdf}
\usepackage{textcomp}
\usepackage{url}
\usepackage{footnote}
\newcounter{mytempeqncnt}
\usepackage[singlespacing]{setspace}
\usepackage{etoolbox}
\usepackage{ccaption}
\usepackage[font=footnotesize,labelfont=rm,figurename=Fig.]{caption}
\IEEEoverridecommandlockouts

\usepackage[table]{xcolor}	

\newcounter{inlineequation}
\DeclareMathAlphabet{\mathpzc}{OT1}{pzc}{m}{it}
\definecolor{Gray}{gray}{0.9}
\definecolor{LightCyan}{rgb}{0.88,1,1}
\usepackage{tcolorbox}
\usepackage{tabularx}
\usepackage{array}
\tcbuselibrary{skins}

\newcolumntype{Y}{>{\raggedleft\arraybackslash}X}

\tcbset{tab1/.style={fonttitle=\bfseries\large, fontupper=\normalsize\small, colupper=black!40!black,
colback=white!10!white,colframe=black!75!black,colbacktitle=black!40!black,
coltitle=black,center title,freelance,frame code={
\foreach \n in {north east,north west,south east,south west}
;}}}

\tcbset{tab2/.style={fonttitle=\bfseries, colupper=black!40!black,
colback=white!10!white,colframe=black!50!black,colbacktitle=black!40!black,
coltitle=black,center title},height = 6.0cm, box align=top,before=,after=\hfill
}

\tcbset{tab3/.style={enhanced,fonttitle=\bfseries,fontupper=\normalsize\small,colupper=black!40!black,
colback=white!10!white,colframe=black!50!black,colbacktitle=black!30!black,
coltitle=black,center title}}

\tcbset{tab4/.style={fonttitle=\bfseries\small,fontupper=\normalsize\small,colupper=black!10!black,
colback=white!10!white,colframe=black!10!black,colbacktitle=white!10!white,
coltitle=black,center title}}

\tcbset{tab5/.style={fonttitle=\bfseries\small,fontupper=\normalsize\small,colupper=black!10!black,
colback=white!10!white,colframe=black!10!black,colbacktitle=white!10!white,
coltitle=black,center title}}

\begin{document}

\title{Spatial and Social Paradigms for Interference and Coverage Analysis in Underlay D2D Network}
\author{\IEEEauthorblockN{Hafiz Attaul Mustafa\IEEEauthorblockA{$^1$}, Muhammad Zeeshan Shakir\IEEEauthorblockA{$^2$}, Muhammad Ali Imran\IEEEauthorblockA{$^3$}, and Rahim Tafazolli\IEEEauthorblockA{$^1$}}\\
\vspace{0.05in}
\fontsize{11}{11}\selectfont
\IEEEauthorblockA{$^1$Institute for Communication Systems, University of Surrey, Guildford, UK.}\\
\vspace{0.05in}
\fontsize{11}{11}\selectfont\ttfamily\upshape
\{h.mustafa, r.tafazolli\}@surrey.ac.uk \\
\vspace{0.05in}
\fontsize{11}{11}\selectfont\rmfamily\
\IEEEauthorblockA{$^2$School of Engineering and Computing, University of the West of Scotland, Paisley, UK.}\\
\vspace{0.05in}
\fontsize{11}{11}\selectfont\ttfamily\upshape
Muhammad.Shakir@uws.ac.uk \\
\vspace{0.05in}
\fontsize{11}{11}\selectfont\rmfamily\
\IEEEauthorblockA{$^3$School of Engineering, University of Glasgow, Glasgow, UK.}\\
\vspace{0.05in}
\fontsize{11}{11}\selectfont\ttfamily\upshape
Muhammad.Imran@glasgow.ac.uk \\

\thanks{Copyright (c) 2015 IEEE. Personal use of this material is permitted. However, permission to use this material for any other purposes must be obtained from the IEEE by sending a request to pubs-permissions@ieee.org.}
}
\maketitle
\begin{abstract}
The homogeneous Poisson point process (PPP) is widely used to model spatial distribution of base stations and mobile terminals. The same process can be used to model underlay device-to-device (D2D) network, however, neglecting homophilic relation for D2D pairing presents underestimated system insights. In this paper, we model both spatial and social distributions of interfering D2D nodes as proximity based independently marked homogeneous Poisson point process. The proximity considers physical distance between D2D nodes whereas social relationship is modeled as Zipf based marks. We apply these two paradigms to analyze the effect of interference on coverage probability of distance-proportional power-controlled cellular user.  Effectively, we apply two type of functional mappings (physical distance, social marks) to Laplace functional of PPP. The resulting coverage probability has no closed-form expression, however for a subset of social marks, the mark summation converges to digamma and polygamma functions. This subset constitutes the upper and lower bounds on coverage probability. We present numerical evaluation of these bounds on coverage probability by varying number of different parameters. The results show that by imparting simple power control on cellular user, ultra-dense underlay D2D network can be realized without compromising the coverage probability of cellular user.
\end{abstract}

\begin{IEEEkeywords}
 D2D communication; IMPPP, Zipf marks; social relation; coverage probability.
\end{IEEEkeywords}

\section{Introduction}\label{intro}
\IEEEPARstart{T}{o} meet capacity demands of cellular networks, the blanket approach calls for cost-effective solution of large number of small cell deployments \cite{Huawei_2014, Qualcomm_2014}. However, this approach incurs capital and operational expenditure (CAPEX/OPEX). To eliminate these costs, device-to-device (D2D) communication, as an underlay network, can be considered as one promising solution \cite{5350367, 6163598, 6957148}. Since mobile terminals are battery operated, and hence constrained by limited power, the underlay D2D communication is intuitively best suitable for proximity services such as content sharing, social networking etc. Such type of proximity based D2D communication has been standardized in Third Generation Partnership Project (3GPP) Long Term Evolution (LTE) Release-12 \cite{3gpp_36.843, 3gpp_22.803, 6807945}.

The D2D underlay network offers increased area spectral efficiency, higher capacity, better coverage, very low end-to-end latency, however, this coexistence poses challenging interference management due to existence of intra-cell cross-tier interference. To analyze such interference, stochastic geometry is a valuable tool. The proximity based D2D communication can be modeled by homogeneous Poisson point process (PPP). However, considering only spatial distribution results in underestimated system insights. For more accurate analysis, we associate social relationship with spatial distribution. In this context, the opportunistic type of D2D communication can be considered from two perspectives, 1). spatial proximity, and 2). social relationship. The first considers channel effects (e.g., physical distance, channel conditions, path-loss etc.) whereas second is more generic to cover any type of homophilic relationship such as social ties (friend, family, colleague, co-author etc.) or inter-dependency bonds (physical contacts, financial exchanges, commodity trades, group participation etc.) \cite{6070945}. The diverse nature of homophilic relations turns the modeling problem intractable. Therefore, we consider one type of homophilic relation between D2D nodes i.e., common content requests of popular files.
\subsection{Related Work}\label{relwork}
The stochastic geometry\footnote{The stochastic framework provides only statistical guidance on performance analysis. This is because it considers probabilistic measure and does not consider instantaneous effects. Though the framework does not provide microscopic view, it provides macroscopic level performance analysis. In industry, the feasibility of proposed scenario/solution can be verified at macroscopic level by stochastic analysis. If the performance does not lie within statistical upper and lower bounds, the solution can be rejected immediately. However, if performance indicators lie within bounds, further sophisticated approaches (based on instantaneous measurements) can be adopted for fine tuning.} is widely used to analyze cellular networks. Specifically, homogeneous PPP and its variant marked PPP (MPPP) are used to model spatial distribution of macro cells, small cells, and cellular users \cite{6515339, 6620410}. In \cite{6515339}, the interference locations in down-link channel of multi-cell heterogeneous network are modeled as MPPP where the marks correspond to the channel (small-scale and large-scale fading) between the interferers and target receiver. Similar to this, in \cite{6620410}, the propagation-loss model of down-link channel of base stations (BSs), enriched by independent exponential marking, is modeled as independently marked PPP (IMPPP).

In case of D2D network, these tools are used in \cite{xu_transmission_2013, 7056528, 6953066, 6504260} to derive the analytic expression for transmission capacity and outage probability. In \cite{7063541}, the authors assume homogeneous PPP (to model spatial distribution) and Zipf distribution (to model demands/requests of popular contents) and derive probability of successful content delivery in the presence of D2D interference. However, in these papers, no notion of social and spatial constraints on interfering D2D nodes are considered to analyze interference effect on cellular user. The results of \cite{6504260} are based on MPPP for D2D user density, however, similar to the previous papers, no analytic representation of some criterion-based D2D pairing has been considered.

In \cite{7421348}, the authors derive coverage probability of D2D receiver to receive portion of the file. The locations are modeled by homogeneous PPP. This 2-file distributed caching system considers simple fixed caching probability without considering Zipf distribution of files. This paper is related to our current work, however, we do not focus on caching application and provide coverage probability of cellular user to meet certain threshold. Moreover, we consider $N$-file Zipf distribution to associate marks to each user location.

In this paper, we extend our previous work \cite{7164238,7145780} and provide novel two paradigm approach to analyze coverage probability of cellular user in underlay D2D communication. We consider physical distance based D2D pairing since it has major contributions towards channel effects. For the social relation, we consider Zipf based common popular file requests by D2D nodes. The two constraints (spatial and social) results into intractable Laplace functional. To approximate coverage probability of cellular user, we derive upper and lower bounds by considering two practical values of shape parameter of Zipf distribution. To best knowledge of authors, such type of joint spatial and social analysis for underlay D2D network is not present in the literature.
\subsection{Contributions}\label{contributions}
\begin{enumerate}
\item We model proximity and social relationship of interfering D2D nodes as proximity based independently marked homogeneous Poisson point process (pIMPPP) to characterize interference impact on coverage probability of cellular user.
\item Based on pIMPPP, we derive analytic expressions for average coverage probability.
\item We derive upper and lower bounds, in the form of digamma and polygamma functions, on coverage probability for two values of shape parameter of Zipf distribution.
\item By numerical evaluations, we show that simple distance-proportional power control on cellular user can be chosen to enhance capacity of the network (user density).
\end{enumerate}
\subsection{Organization}
The rest of the paper is organized as follows: In Section II, we present system model of cellular network with underlay D2D communication. In Section III, we derive analytic expressions for coverage probability of cellular user based on pIMPPP distributed D2D nodes. The upper and lower bounds on coverage probability are presented for two practical values of shape parameter of Zipf distribution. In Section IV, we present numerical evaluations by varying different parameters such as user density, D2D pairing distance, power control of cellular user, target threshold, transmit power of D2D pairs, and path-loss exponent. Conclusion is drawn in Section V.
\section{System Model}\label{sys_mod}
\begin{figure}[t]
\centering
\includegraphics[width = 3.25 in, height = 2 in]{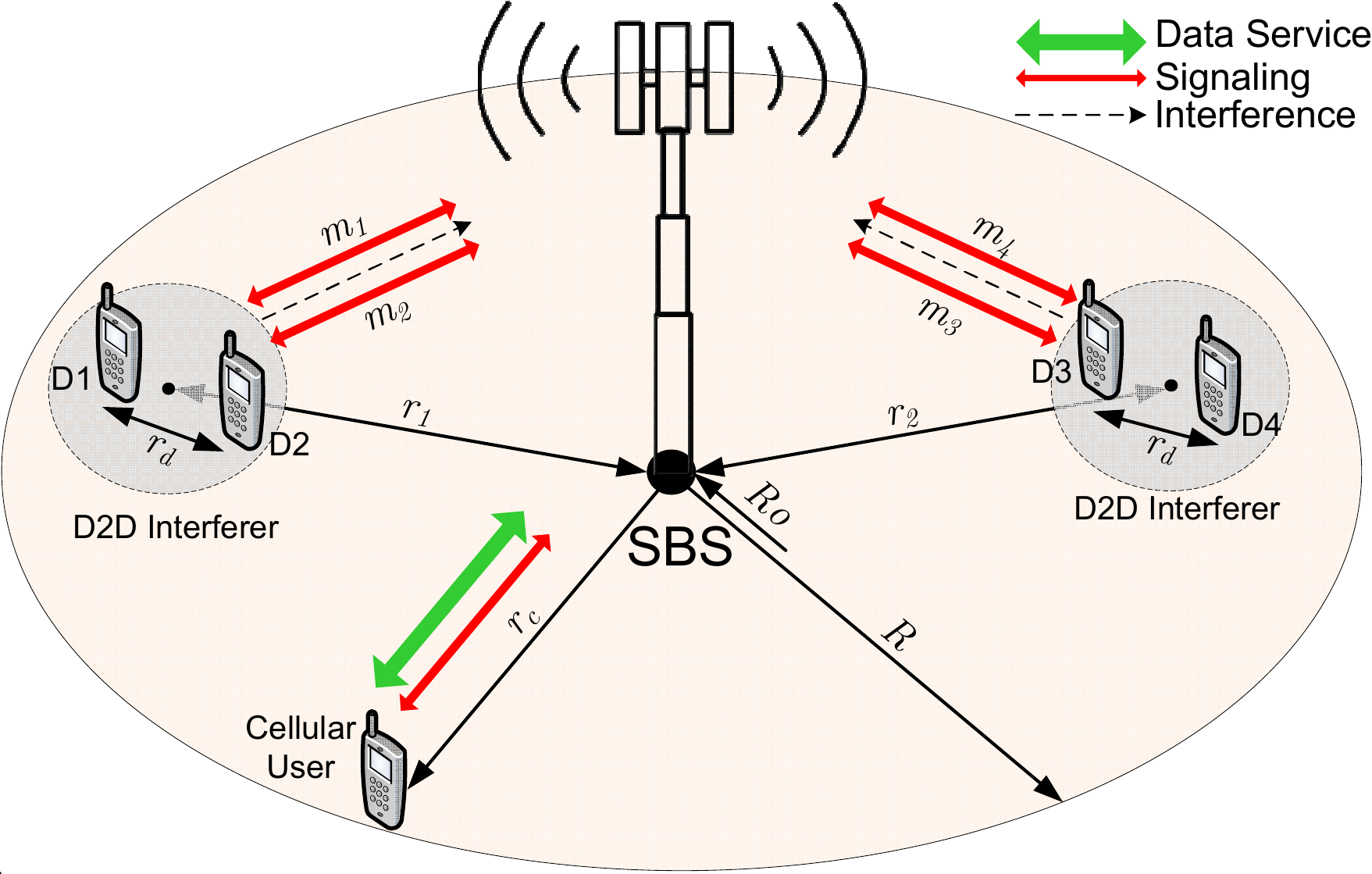}
\caption{System Model for Interference Analysis}\label{Figure:sm}
\vspace{-5mm}
\end{figure}

We consider single cell scenario comprising small cell BS (SBS), cellular user and potential D2D users as shown in Fig. \ref{Figure:sm}.
The up-link channel of cellular user is shared by potential D2D users in time division duplex (TDD) mode, hence causing accumulated interference. The D2D pairs are network assisted (only control/signaling information) whereas cellular user is provided both control and data services.

In the following, we define PPP and its variants to understand spatial and social distribution of potential D2D users.
\begin{defn}\label{def1}
For bounded and mutually disjoint sets $A_i$ from $i = 1,...,k$, the homogeneous PPP is defined by its finite dimensional distributions \cite{blaszczyszyn_stochastic_2009}:
\begin{align}
\mathbb{P}\{\Phi(A_1)=n_1,...,\Phi(A_k) = n_k\} = \prod_{i=1}^{k} \bigg(\frac{\exp(-\lambda A_i) \lambda A^{n_i}_i}{(n_i)^!}\bigg), \IEEEnonumber
\end{align}
where $\lambda$ is the intensity parameter. For each point $x_i$, homogeneous PPP can be represented as $\Phi = \{x_i\}$.
\end{defn}
\begin{defn}\label{def2}
An MPPP is homogeneous PPP with point vector in $\mathbb{R}^d$ and random mark vector in $\mathbb{R}^l$ attached to each point \cite{blaszczyszyn_stochastic_2009}. It can be represented as $\Phi=\{x_i,m_i\}$, where $x_i$ are points and $m_i$ are associated marks. If the points and marks are independent from each other, it is known as IMPPP process represented as $\tilde{\Phi}=\{x_i,m_i\}$.
\end{defn}
\begin{defn}\label{def3}
The pIMPPP is defined as IMPPP process with function mapping  $p: \mathbb{R}^2 \mapsto [0, 1]$ to perform spatial proximity based thinning. This process can be represented as $\tilde{\Phi}^p = p(\{x_i,m_i\})$.
\end{defn}
We assume that the cellular user is uniformly distributed whereas potential D2D users are distributed according to homogeneous PPP $\Phi=\{x_i\}$ where $x_i$'s are locations of each point on measurable space $\mathbb{R}^2$. For realistic interference analysis, we associate marks $m_i \in \mathbb{N}$ to each point $x_i$. The resulting process is MPPP. The marks are based on common content requests which follow Zipf distribution. Since, up-link request of each D2D user is associated with that user only and is independent of locations of other D2D users, the point process forms IMPPP process $\tilde{\Phi}=\{x_i,m_i\}$.

The cellular and potential D2D users are distributed in the coverage area bounded between SBS radius $R$ and the protection region $R_0$. The distance between SBS and cellular user is $r_c$ which is used to calculate path-loss. Every successful D2D pair has a distance of $r_d$ between nodes. The distance between D2D interferer and SBS is $r_i$. To simplify the analysis, we assume opportunistic D2D communication without considering quality of service for successful D2D pair and hence do not take into account the distance between D2D receiver and cellular user\footnote{The distance between D2D receiver and cellular user is random for whole set of underlay D2D nodes. It has well known distribution \cite{moltchanov2012distance1}, however, we intend to provide this type of analysis in extended work in future.}. The channel model assumes distance dependent path-loss and Rayleigh fading. The simple singular path-loss model\footnote{More complex models for path-loss and channel gains\cite{6042301,5226963} can be assumed, however, they result in decreased tractability, and are left to the future work. In this research, our approach is to assume standard models and analyze more realistic stochastic models for spatial distribution and D2D pairing.} $({r_c}^{-\alpha})$ is assumed where the protection region ensures the convergence of the model by avoiding $r_c$ to lie at the origin. The radius $R_0$ is very small i.e., $R_0 \ll R$ such that it can be considered as an atom in point process terminology i.e., $R_0 \sim 0$. The power, associated with channel gain, follows exponential distribution with mean $\mu$. The distance $r_c$ follows uniform probability distribution function (pdf) as follows \cite{6953066}:
\begin{align}
f(r_c)=\frac{2r_c}{R^2},f({\theta})=\frac{1}{2\pi},
\label{frc}
\end{align}
where $R_0\leq r_c \leq R$ and $0 \leq \theta \leq 2\pi$.
\begin{figure}[t]
\centering
\includegraphics[width=3.25 in, height = 2.2in]{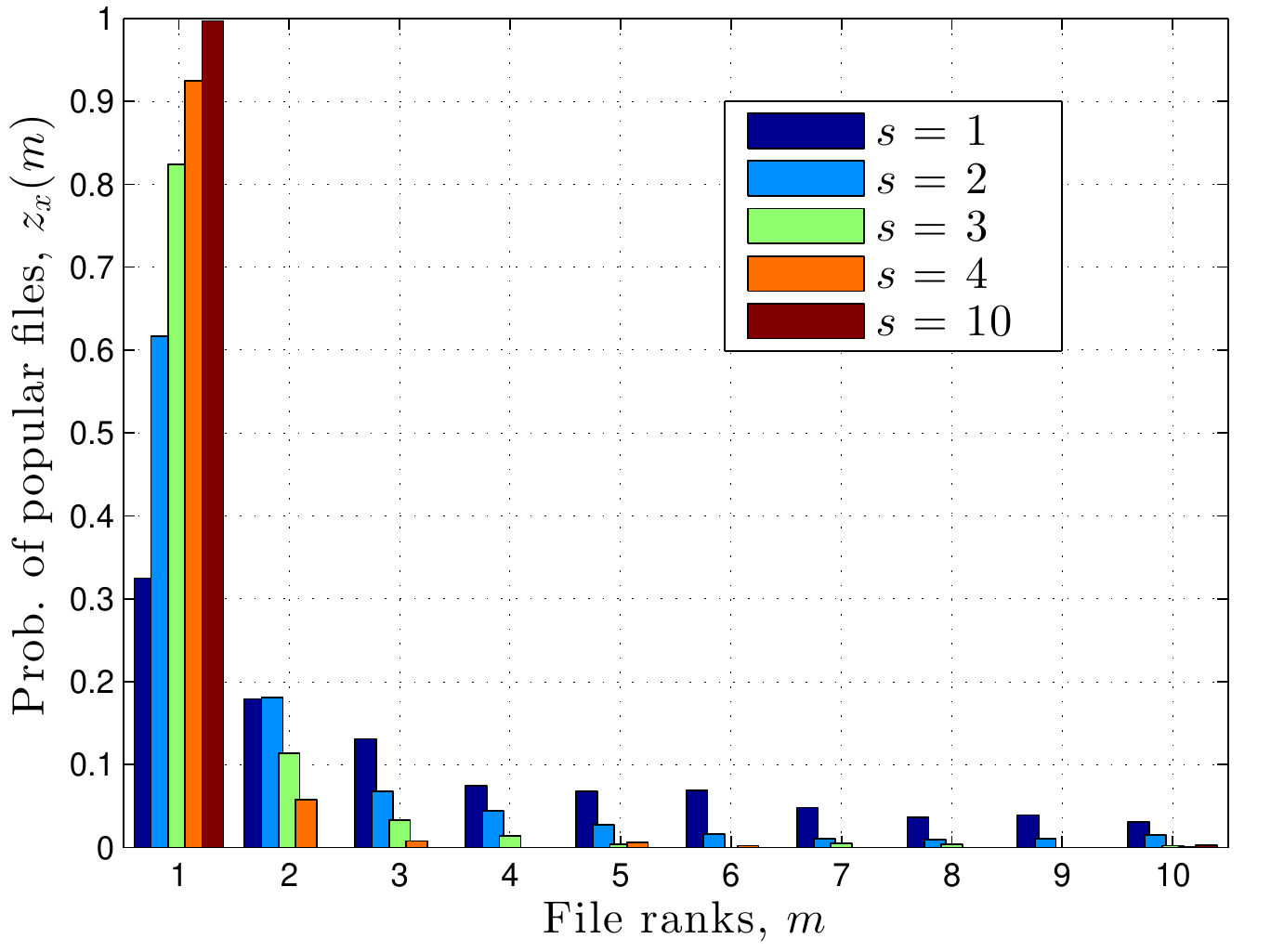}
\caption{Zipf based marks of IMPPP for different values of shape parameter $s$.}\label{Figure:zipf_marks}
\vspace{-5mm}
\end{figure}
\section{Coverage Probability of Cellular User}\label{retprob}
In this section, we derive coverage probability of cellular user by considering not only the distance constraint but also the popularity of requested contents.
\subsection{Social Relation}
The Zipf distribution models the popularity of files in the form of ranks. The probability mass function of Zipf distribution is given as \cite{devroye_non-uniform_2013}:
\begin{align}
z_x(k) \stackrel{(a)}{=}& \frac{1}{k^s \displaystyle \sum_{i = 1}^{\infty} (1/i)^s} = \frac{1}{k^s \zeta(s)},	\IEEEnonumber
\\ \stackrel{(b)}{=}& \frac{1}{k^s \displaystyle \sum_{i = 1}^N (1/i)^s} = \frac{1}{k^s H_{N,s}},
\label{Zipfpmf}
\end{align}
where $(a)$ involves Riemann zeta function and $(b)$ follows the fact $\displaystyle \lim_{N\rightarrow \infty} H_{N,s}=\zeta(s)$ \cite{kiskani_capacity_2016}. Since our analysis is based on $N$ popular files, associated with location $x$, we will consider $(b)$ throughout this paper. In Equ. (\ref{Zipfpmf}), $k$ is the rank of the file, $s$ is the tail index or shape parameter and $H_{N,s}$ is the $N^{th}$ generalized harmonic number.

The Zipf based marks associated with each D2D user are shown in Fig. \ref{Figure:zipf_marks}. The shape parameter $s$ controls the popularity of files. The most popular contents have higher probability of request by potential D2D users. On contrary, the less popular contents will result in lower probability of D2D pairing and hence less interference.  For example, the chances of rank 1 ($m=1$) file with $s=10$ are 99.9\%. The higher value of $s$ converges the popularity of files towards lower ranks. The popular contents follow Zipf's Law and hence, the interference due to potential D2D users is controlled by power law.
\subsection{Spatial Proximity}\label{phyrel}
In order to consider proximity for D2D pairing along with Zipf based associated marks, we assume reduced path-loss (shortest distance) between potential D2D users. All nodes that do not meet this criterion are excluded from D2D pairing. In this context, we convert IMPPP process $\tilde{\Phi}$ to pIMPPP $\tilde{\Phi}^p$ where $p: \mathbb{R}^2 \mapsto [0, 1]$ performs thinning to analytically capture the shortest distance based selection of points, for D2D pairing, that have higher marks for content sharing applications.

We introduce retention $p(r_d)$ as the probability that all points with distance ($\leq$) $r_d$ are considered for D2D pairing. The probability $p(r_d)$ is given as \cite{7164238,7145780}:
\begin{align}
p(r_d) = p = \, 1 - \exp(-\pi \lambda r_{d}^{2}).
\label{prd}
\end{align}

The locations of potential D2D users constitute measurable space in $\mathbb{R}^2$ and associated marks are mutually independent random vectors in $\mathbb{N}$. By incorporating retention probability (\ref{prd}), the Laplace functional of pIMPPP is given as \cite{blaszczyszyn_stochastic_2009}:
\begin{align}
\lefteqn{\mathcal{L}_{\tilde{\Phi}^p} (f)}	\IEEEnonumber
\\ = & \exp\bigg\{- \int_{\mathbb{R}^2} \bigg[1-\sum_{m=1}^N \exp\bigg(-f(x,m)z_{x}(m)\bigg)\bigg] p\: \lambda\: dx\bigg\},	\IEEEnonumber
\end{align}
where $f(x,m)$ is the real function defined on $\mathbb{R}^2$ and $\lambda$ is the intensity measure of points (in this case D2D user density).
\subsection{Illustrative Example}\label{phyrel}
\begin{figure}[t]
\centering
\includegraphics[width=3.25 in]{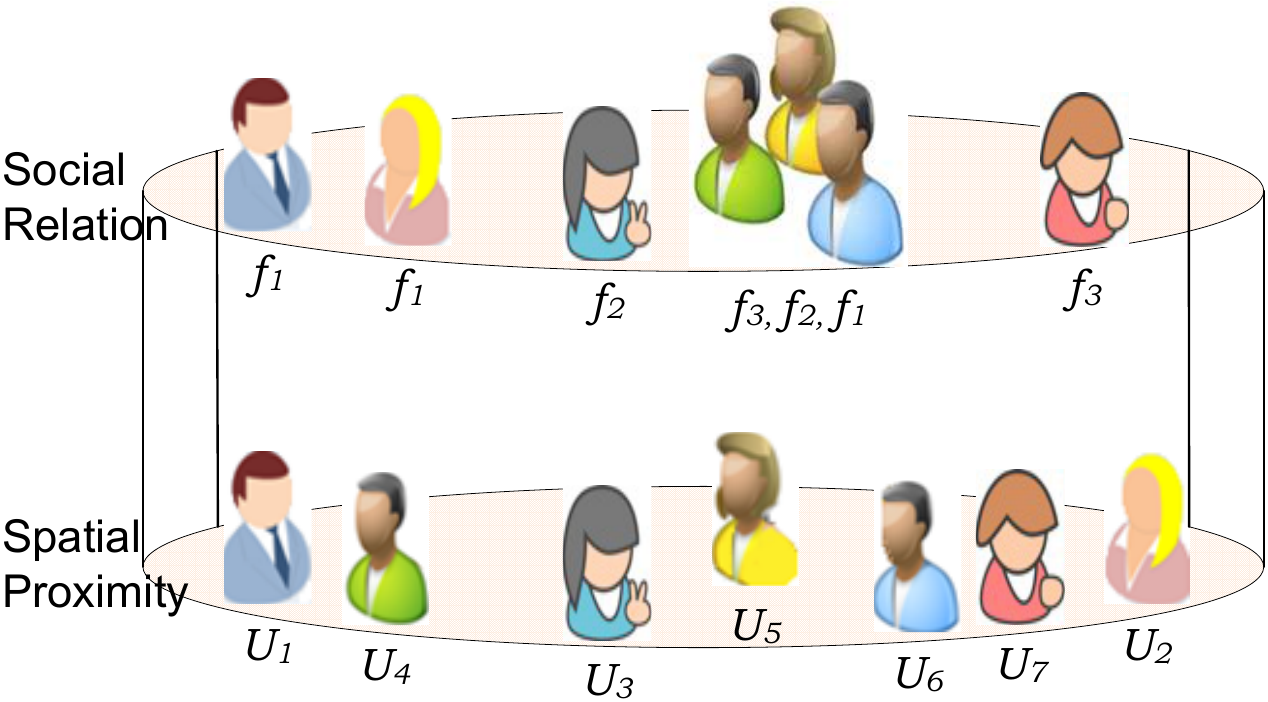}
\caption{Social and spatial relation between users/locations for D2D pairing.}\label{Figure:socsparel}
\vspace{-3mm}
\end{figure}

The social and spatial relations play key role to realize D2D pairs. These relations are diverse in nature. However, as mentioned in Sec. \ref{intro}, in this research we consider physical distance based spatial proximity and common request based social relation. For illustration purpose, we show these two relations in Fig. \ref{Figure:socsparel}.
In this figure, we can see that each user has different social and spatial relation with every other user in the coverage area. For example, $U_1$ is physically away from $U_2$, however, socially it is well connected i.e., requesting the most popular common file $f_1$ where subscript $1$ shows the rank of the file (See Fig. \ref{Figure:zipf_marks}). Similarly, $U_2$ has strong physical relation with $U_7$ but socially not well connected (i.e., $U_2$ requests $f_1$ whereas $U_7$ requests $f_3$). The user $U_3$ is spatially and socially closer to $U_5$ (both request $f_2$) and hence qualify for D2D pairing. In this research, we model social relation of users by multivariate Zipf distribution (Equ. \ref{Zipfpmf}) and physical relation by nearest neighbor distribution function (Equ. \ref{prd}).
\subsection{Coverage Probability}\label{covprob}
In order to derive average coverage probability of cellular user, we assume interference-limited environment ($\sigma^2=0$). In the following coverage analysis, we consider signal-to-interference ratio (SIR) instead of signal-to-interference-and-noise-ratio (SINR). Due to large number of underlay D2D users, we assume interference dominated regime where noise variance $\sigma^2$ can be ignored\footnote{In LTE-based OFDMA systems, the value of $\sigma^2$ depends on Boltzmann constant, absolute temperature, number of carriers $N_C$, allocated bandwidth ($B$) and total bandwidth ($B_T$) of the system. In case of $B_T$ = 10 MHz, $N_C$ = 600, $B$ between [0.5 5] MHz (as the up-link shared bandwidth by underlay D2D network), the value of $\sigma^2$ is between [$1.8 \times 10^{-15}$ $1.8 \times 10^{-14}$] W. Since, we have negative sign with $\sigma^2$, to observe and maximize its effect, we choose minimum value of $\tau$ (e.g., -5 dB), minimum value of $r_c^{\alpha}$ (e.g., 10 m from BS), and maximum value of $p_c$ (e.g., 20 dBm). For these values, noise contribution is $\exp\big(-\sigma^2\tau p^{-1}_{c} r^{\alpha}_c\big) = 0.999$ which has no impact on overall analysis.}. In this case, SIR is given as:
\begin{equation*}
\textnormal{SIR}_{SBS}  =\frac{ p_c f_{c} r^{\alpha(\epsilon -1)}_{c}}{\displaystyle \sum_{i \in \tilde{\Phi}} p_{i} f_{i} r^{-\alpha}_{i} m^{-1}_i },
\end{equation*}
where subscripts $\{_c, \:_i\}$ are used for cellular user and D2D interferers, respectively. The $p_c$ and $p_i$ are transmit powers; the $f_{c}$ and $f_{i}$ are Rayleigh based small-scale fading where power gain follows exponential distribution with mean $\mu$; the distance dependent path-losses are $r^{-\alpha}_{c}$, and $r^{-\alpha}_{i}$. We assume distance-proportional fractional power control $r^{\alpha \epsilon_{(\cdot)}}_{(\cdot)}$ \big($\epsilon_{(\cdot)} \in [0,1]$\big) for all users. However, it does not make sense to invert path-loss of the interfering channel by power control. Therefore, we assume power control for cellular user and no power control for potential D2D interferers. To allow underlay D2D communication, the SBS should switch-off power control that converts $r^{\alpha(\epsilon_{i}-1)}_{i}=r^{-\alpha}_i$ to ensure better coverage of the cellular user with power control $r^{\alpha(\epsilon-1)}_{c}$. The marks $m_i$ follows Zipf law to model the up-link request of popular files.
\setcounter{mytempeqncnt}{\value{equation}}
\setcounter{equation}{9}
\begin{table*}[!htb]
\makeatletter
\newcommand*{\compress}{\@minipagetrue}
\makeatother
\renewcommand{\arraystretch}{1.25}
\caption{Closed-form Expressions for Upper and Lower Bounds on $g(x,r_c)$.}\label{Table:upperlowerbounds}
\vspace{1mm}
\centering
\begin{tcolorbox}[tab2,tabularx={>{\raggedright\arraybackslash}p{0.65in}|>{\raggedright\arraybackslash}p{0.8in}|X}]
\textbf{Shape Parameter}		& \textbf{Constants}	& \textbf{Function $g(x,r_c)$}	\\ \hline 
\begin{equation*}s = 1\end{equation*}		& 	 \begin{equation*} A = \frac{\tau p^{-1}_c p_i}{\mu}\end{equation*}		&	\begin{equation}
g^{UB}(x,r_c) = \frac{1}{H_N}\bigg[\Psi\bigg(N+1+\frac{A r^{\alpha(1-\epsilon)}_c}{x^{\alpha}}\bigg)-\Psi\bigg(1+\frac{A r^{\alpha(1-\epsilon)}_c}{x^{\alpha}}\bigg)\bigg] \label{lowerbound}\end{equation}	\\ \hline 
\begin{equation*}s = 10\end{equation*}			&	\begin{eqnarray*}B_{10} = \frac{1}{93555},\\ C_{10} = \frac{1}{362880}\end{eqnarray*} & \begin{equation} g^{LB}(x,r_c) = \frac{1}{\bigg({1+\frac{A r^{\alpha(1-\epsilon)}_c}{x^{\alpha}}}\bigg)H_{N,s}}\bigg[B_{10} \pi^s-C_{10}\Psi(s-1, N+1)\bigg]\label{upperbound}\end{equation} \\ \hline
\multicolumn{3}{l}{\multirow{1}{*}{$\Psi(\cdot)$ is digamma and $\Psi(n, \cdot)$ is the $n^{th}$ polygamma function, respectively.}}
\end{tcolorbox}
\vspace{-4mm}
\end{table*}

The average coverage probability of cellular user is:
\setcounter{equation}{3}
\begin{align}
p_{cov}^{c} = & \,\mathbb{E}_{r_{c}}\big[\mathbb{P}[\textnormal{SIR}_{SBS} \geq \tau]\,|\,r_{c}\big], \IEEEnonumber
\\ =& \,\mathbb{E}_{r_{c}}\big[\mathbb{P}[(f_{c} \geq \frac{\tau I_{A}}{p_{c} r^{\alpha(\epsilon-1)}_{c}})]\,|\,r_{c}\big],	\label{SIR_AppA}
\end{align}
where 
\begin{align}
I_{A} = \sum_{i \in \tilde{\Phi}} p_{i} f_{i} r^{-\alpha}_{i} m^{-1}_i,
\label{IA}
\end{align}
is the interference due to D2D users in the coverage area.

In (\ref{SIR_AppA}), the coverage probability depends on $f_{c}, r^{-\alpha}_{c}, f_{i}, r^{-\alpha}_{i}$. Conditioning on $g=\{f_{i}, m_i\}$, the coverage probability of cellular user for a given transmit power $p_c$ is
\begin{align}
\mathbb{P}[\textnormal{SIR}_{SBS} \geq \tau]\,|\,r_{c},g =& \, \int_{x=\frac{\tau I_{A}}{p_{c} r^{\alpha(\epsilon-1)}_{c}}}^\infty \exp(- x)dx,	\IEEEnonumber
\\=& \, \exp\bigg(-\tau p^{-1}_{c} r^{\alpha(1-\epsilon)}_{c} I_{A}\bigg).
\label{SIR_AppA1}
\end{align}
De-conditioning by $g$, (\ref{SIR_AppA1}) results into:
\begin{align}
\mathbb{P}[\textnormal{SIR}_{SBS} \geq \tau]\,|\,r_{c} =& \, \mathbb{E}_{g}\bigg[\exp\big(-\tau p^{-1}_{c} r^{\alpha(1-\epsilon)}_{c} I_{A}\big)\bigg],	\IEEEnonumber
\\ =& \, \mathbb{E}_{g}\bigg[\exp\big(- s_{c} I_{A}\big)\bigg],	\IEEEnonumber
\\ =& \,\mathcal{L}_{I_{A}}\big(s_{c}\big),
\label{SIR_AppA2}
\end{align}
where $s_{c}=\tau p^{-1}_{c} r^{\alpha(1-\epsilon)}_{c}$. Putting the value of $I_{A}$ from (\ref{IA}) into (\ref{SIR_AppA2})
\begin{align}
\mathcal{L}_{I_{A}}\big(s_{c}\big) = & \,\mathbb{E}_{\tilde{\Phi},f_{i}}\bigg[ \exp\bigg(-s_{c} \displaystyle \sum_{i \in \tilde{\Phi}} p_{i} f_{i} r^{-\alpha}_{i} m^{-1}_i\bigg)\bigg],	\IEEEnonumber
\\  = & \,\mathbb{E}_{\tilde{\Phi},f_{i}}\bigg[\prod_{i \in \tilde{\Phi}} \exp\bigg(-s_{c} p_{i} f_{i} r^{-\alpha}_{i} m^{-1}_i\bigg)\bigg],	\IEEEnonumber
\\ = & \,\mathbb{E}_{\tilde{\Phi}}\bigg[\prod_{i \in \tilde{\Phi}} \mathbb{E}_{f_{i}} \big[\exp(-s_{c}  p_{i} f_{i} r^{-\alpha}_{i} m^{-1}_i)\big]\bigg],	\IEEEnonumber
\\ = & \,\mathbb{E}_{\tilde{\Phi}}\bigg[\prod_{i \in \tilde{\Phi}} \mathbb{E}_{p^{\alpha \epsilon}_{i}} \bigg(\frac{1}{1+s_{c}  p_i r^{-\alpha}_{i} m_i}\bigg)\bigg],	\IEEEnonumber
\\ = & \,\mathbb{E}_{\tilde{\Phi}}\bigg[\prod_{i \in \tilde{\Phi}} \underbrace{\bigg(\frac{\mu}{\mu +s_{c} p_{i} r^{-\alpha}_{i} m^{-1}_i}\bigg)}_{f(x,m)}\bigg],
\label{SIR_AppA3}
\end{align}
where (\ref{SIR_AppA3}) results from the i.i.d distributions of $f_{i}$ and further independence from the underlay IMPPP process.

The probability generating functional for a function $f(x,m)$ with retention $p$ from (\ref{prd}) implies:
\begin{align}
\lefteqn{\mathcal{L}_{I_{A}}\big(s_{c}\big)}		\IEEEnonumber
\\ &= \exp\bigg(- \int_{\mathbb{R}^2} \bigg[1-\displaystyle \sum_{m=1}^N f(x,m)  z_{x}(m)\bigg] p\: \lambda\: dx\bigg),	\IEEEnonumber
\\ &= \exp\bigg(-\int_{\mathbb{R}^2} \bigg[1- \displaystyle \sum_{m=1}^N f(x,m) \frac{m^{-s}}{H_{N,s}}\bigg] p\: \lambda\: dx\bigg),\IEEEnonumber
\\ &= \exp\bigg(-2\pi \lambda\, p \int^{\infty}_{R_0} \bigg[1- \underbrace{\sum_{m=1}^N \frac{x^{\alpha} \mu m^{(1-s)}}{(x^{\alpha} \mu m + s_{c} p_{i})H_{N,s}}}_{g(x,r_c)} \bigg] x dx\bigg).
\label{PGFL_AppA}
\end{align}

The function $g(x,r_c)$ in (\ref{PGFL_AppA}) cannot be solved in closed-form in general, however, we can find upper and lower bounds for special cases of $s \in \mathbb{N}$. These bounds subsequently allows upper and lower bounds on coverage probability of cellular user. For example, for a given population of file ranks, the lower bound on $g(x,r_c)$ is controlled by minimum value of $s\in \mathbb{N}$ (i.e., $s=1$) which shows minimum D2D pairs and puts upper bound on coverage probability of cellular user. Similarly, upper bound is controlled by maximum value of $s \in \mathbb{N}$ which can be as high as $\infty$. In order to find practical highest value of $s$, we consider the value where the probability of most popular files reaches 1. This allows maximum D2D pairs and puts lower bound on coverage probability of cellular user. The maximum value of shape parameter is found by increasing it gradually. By increasing shape parameter from $s=1$ to $s=2$, the probability of most popular files approximately doubles (from 0.32 to 0.61). From $s=2$ to $s=4$, it increases by 1.5 times (from 0.61 to 0.92). Further increase in $s$, e.g., $s=10$, increases the probability of popular files to 0.999. If we further increase the value of $s$, the probability of popular files will not increase significantly. Moreover, higher values of $s$ incur higher cost of power function in denominator of Equ. \ref{Zipfpmf}. To reduce computational complexity, we assume the highest value of $s$ as 10 without any significant loss in probability of popular files. For minimum and maximum values of $s$, the function $g(x,r_c)$ in (\ref{PGFL_AppA}) converges to closed-form expressions as shown in Table \ref{Table:upperlowerbounds}. The minimum value of shape parameter i.e., $s=1$ results in lowest probability of popular file requests as can be seen in Fig. \ref{Figure:zipf_marks}. For example, for $s=1$, the probability of file with rank $m=1$ has $32.6\%$ chances of being requested by any potential D2D pair. This minimum value allows closed-form expression for $g(x,r_c)$ \big[Table \ref{Table:upperlowerbounds}, (\ref{lowerbound})\big] and hence, characterizes upper bound on coverage probability of cellular user. Contrary to this, higher value of $s$ means higher probability of popular content requests which results into maximum number of D2D pairs and subsequently more interference on cellular user. For example, for $s=10$, there are $99.9\%$ chances that the file with rank $ =1$ will be requested. This case admits closed-form expression for $g(x,r_c)$ \big[Table \ref{Table:upperlowerbounds}, (\ref{upperbound})\big] and puts lower bound on coverage probability. The resulting average coverage probability of cellular user is given as:
\setcounter{equation}{11}
\begin{align}
p_{cov}^{c} = & \,\mathbb{E}_{r_{c}}\bigg[\exp\bigg(-2\pi \lambda\, p \int^{\infty}_{R_0} \big[1- g(x,r_c)\big] x dx\bigg)\,|r_{c}\bigg], \IEEEnonumber
\\ =& \int^{R}_{R0}\exp\bigg(-2\pi \lambda\, p \int^{\infty}_{R_0} \big[1- g(x,r_c)\big] x dx \bigg)\,\frac{2 r_{c}}{R^{2}} dr_{c}.
\label{SIR_AppA5}
\end{align}

Using (\ref{SIR_AppA5}), the upper and lower bounds on average coverage probability of cellular user takes the following form, respectively:
\begin{align}
\lefteqn{p_{cov}^{c, UB}}	\IEEEnonumber
\\ &= \int^{R}_{R0}\exp\bigg(-2\pi \lambda\, p \int^{\infty}_{R_0} \big[1- g^{UB}(x,r_c)\big] x dx \bigg)\,\frac{2 r_{c}}{R^{2}} dr_{c},
\label{SIR_UB}
\end{align}
\begin{align}
\lefteqn{p_{cov}^{c, LB}}	\IEEEnonumber
\\ &= \int^{R}_{R0}\exp\bigg(-2\pi \lambda\, p \int^{\infty}_{R_0} \big[1- g^{LB}(x,r_c)\big] x dx \bigg)\,\frac{2 r_{c}}{R^{2}} dr_{c}.
\label{SIR_LB}
\end{align}

\begin{figure}[t]
\centering
\includegraphics[width = 1\columnwidth, height = 3 in]{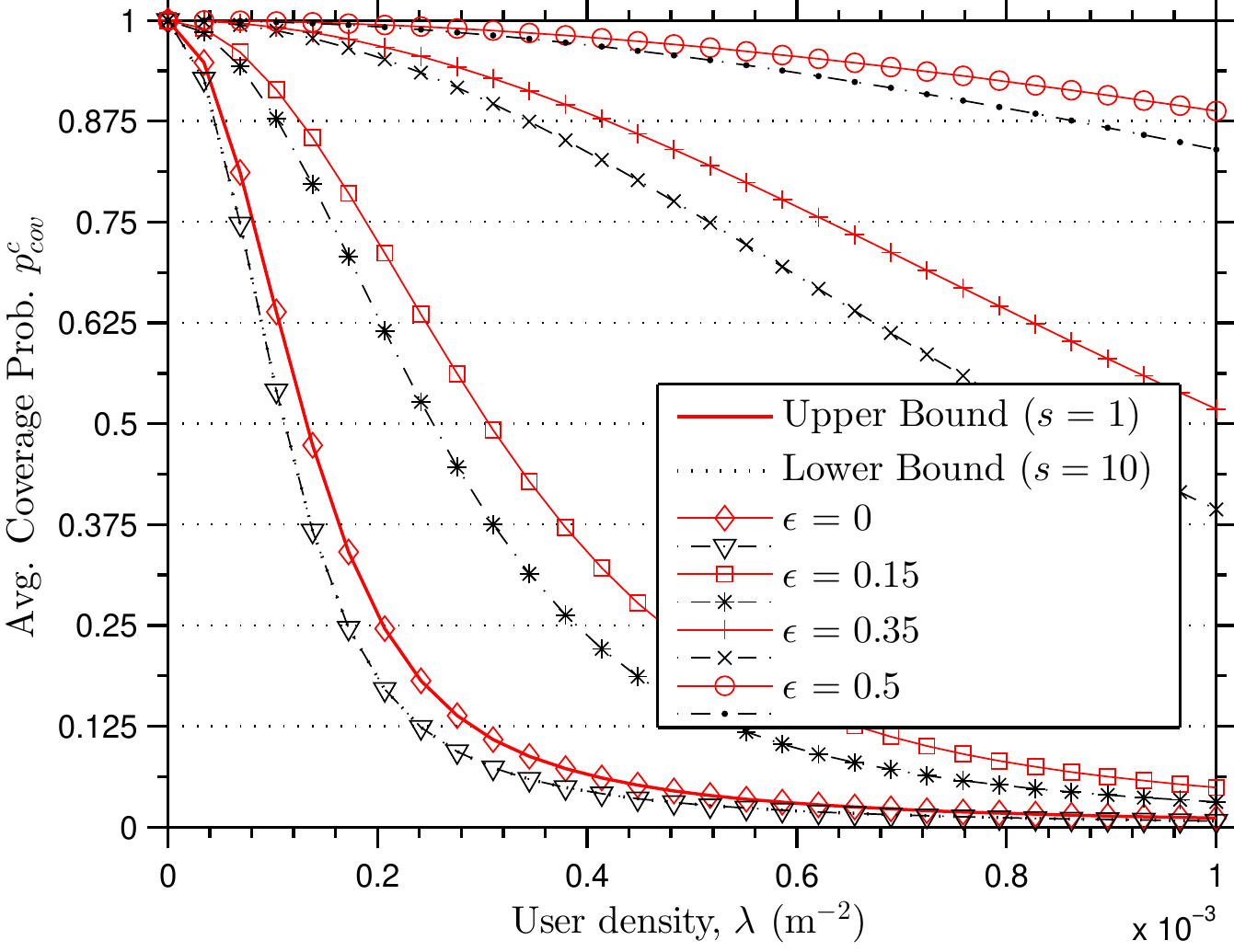}
\caption{Average coverage probability and user density $\lambda$ for $p_{i}=1$ mW, $p_{c}=0.2$ W, $r_d \leq10$ m, and $\tau=15$ dB, $\alpha=4$, and different values of power control factor $\epsilon$.}\label{Figure:covdensityepsilon}
\vspace{-4mm}
\end{figure}
\subsection{Ergodic Rate of D2D Pair}\label{ergodicrate}
The coverage probability of D2D pair is given as \cite{7145780}:
\begin{align}
p_{cov}^{d} = & \, \bigg[\mathcal{L}_{I_{c}}(s_{d}) \, \, \, \mathcal{L}_{I_{A}}(s_{d})\bigg],
\label{SIR_AppB2}
\end{align}
where $I_{c} = p_{c} f_{c} d^{-\alpha}$, $s_{d} = \gamma p^{-1}_{d} r^{\alpha}_{d}$ and $p_d$ is the transmit power of D2D pair.

From \cite[Equ. (17)]{7145780}, the first part of Equ. (\ref{SIR_AppB2}) is approximated as:
\begin{align}
\mathcal{L}_{I_{c}}\big(s_{d}\big) \simeq & \,\frac{1}{1+ (\gamma \frac{p_c }{p_d})^{\frac{2}{\alpha}}\frac{r^2_{d}}{(128 R/45\pi)^2}}.
\label{Lc2}
\end{align}
The second part of Equ. (\ref{SIR_AppB2}) \big(i.e., $\mathcal{L}_{I_{A}}(\cdot)$\big) is given in Equ. (\ref{PGFL_AppA}), however, Laplace functional is calculated at $s_d$ instead of $s_c$ since the reference point is now D2D pair.

Following \cite[Sec. V-A, Equ. (29)]{6953066} and utilizing Equ. (\ref{PGFL_AppA} \& \ref{Lc2}), the ergodic rate of a single D2D pair can be approximated as:
\begin{align}
\mathcal{R}^{d} &= \int_0^{\infty}\frac{1}{1+\gamma} \mathcal{L}_{I_{c}}(s_{d}) \, \, \, \mathcal{L}_{I_{A}}(s_{d}) d\gamma,	\IEEEnonumber
\\&\simeq \int_0^{\infty}\frac{1}{(1+\gamma)\bigg(1+ (\gamma \frac{p_c }{p_d})^{\frac{2}{\alpha}}\frac{r^2_{d}}{(128 R/45\pi)^2}\bigg)} 	\IEEEnonumber
\\ &\quad \exp\bigg(-2\pi \lambda\, p \int^{\infty}_{R_0} \bigg[1- {g(d,r_d)} \bigg] x dx\bigg) d\gamma.
\label{ergrate1}
\end{align}
The lower and upper bounds on $\mathcal{R}^{d}$ can be obtained by considering $A = \frac{\tau p^{-1}_d p_i}{\mu}$ in Table \ref{Table:upperlowerbounds} and replacing $g(d,r_d)$ with Equ. ($\ref{lowerbound}$ \& $\ref{upperbound}$), respectively.
\section{Numerical Results and Discussion}\label{numres}
In this section, we numerically evaluate the analytic expressions of Sec. \ref{retprob} by varying the number of different parameters for a small cell of radius $R=500$ m and a radius of protection region $R_0=1$ m. The distance of cellular user $r_c$ is uniformly distributed. It is integrated over the coverage area to capture average coverage probability of cellular user. The spatial proximity (D2D pairing, $r_d$) is random and reduced path-loss (shortest distance) based D2D pairing is captured by thinning the IMPPP process using retention probability (\ref{prd}). The up-link requests of D2D nodes are modeled by marks where two extreme values of shape parameter of Zipf distribution are considered. These values characterize upper and lower bound on coverage probability of cellular user. We assume $N=10$ popular files to analyze the interference effect of D2D pairs on cellular user due to spatial proximity and social relationship. The solid lines, in all figures, show upper bound ($s=1$) and dotted lines show lower bound ($s=10$) on coverage probability.
\begin{figure}[t]
\centering
\includegraphics[width = 1\columnwidth, height = 3 in]{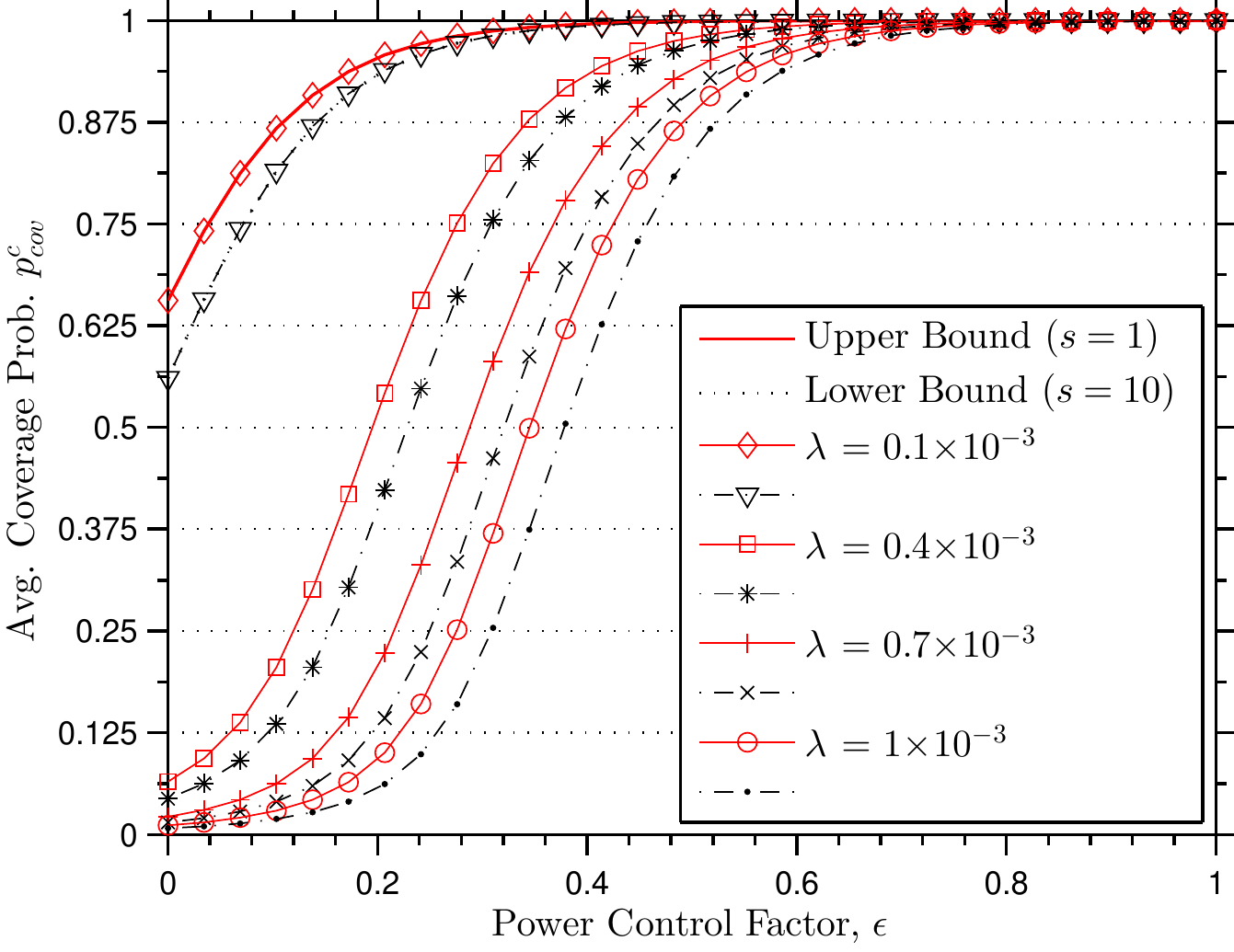}
\caption{Average coverage probability and power control factor $\epsilon$ for $p_{i}=1$ mW, $p_{c}=0.2$ W, $r_d \leq10$ m, and $\tau=15$ dB, $\alpha=4$, and different values of user density $\lambda$.}\label{Figure:fig56_letter}
\end{figure}

The average coverage probability of cellular user in (\ref{SIR_UB}, \ref{SIR_LB}) depends on user density $\lambda$, D2D pair distance $r_d$, SIR threshold of cellular user $\tau$, D2D transmit power $p_i$, transmit power of cellular user $p_c$, path-loss exponent $\alpha$, and power control factor $\epsilon$. The average coverage probability of cellular user for different values of $\lambda$ and $\epsilon$ is plotted in Fig. \ref{Figure:covdensityepsilon}. In this two parameter analysis, there is a monotonically increasing relation between $\lambda$, $\epsilon$ and $p^{c}_{cov}$ as shown in Fig. \ref{Figure:fig56_letter}. Based on this monotonic relation, the value of $\epsilon$ is selected in incremental fashion\footnote{If we consider all other parameters e.g., $p_{i}$, $p_{c}$, $r_d$, $\lambda$ in finding $\epsilon$, then it can be formulated as an optimization problem, however, in present research, the scope is restricted to analyze coverage probability of distance-proportional power-controlled cellular user in underlay D2D network where pairs are made on spatial and social relations.} to show its impact on coverage probability of cellular user. 
The plot in Fig. \ref{Figure:covdensityepsilon} shows that with different power control, the coverage probability of cellular user shows quite different trends. For example, without power control ($\epsilon=0$), the upper bound on coverage probability drops from 0.94 to 0.47 by increasing underlay users from 27 to 108 ($\lambda=0.03\times10^{-3} \rightarrow 0.1\times10^{-3}$). On the other hand, the upper bound with high power control ($\epsilon=0.5$) shows hardly any drop in coverage probability for this range of user density. The power control can be incorporated to realize dense underlay D2D network. The higher power can cause intra-cell (cellular user to D2D receiver) and inter-cell interference (cellular user to cell-edge users of neighboring cells). However, this problem can be solved via optimal power control and resource management.

\begin{figure}[t]
\centering
\includegraphics[width = 1\columnwidth, height = 3 in]{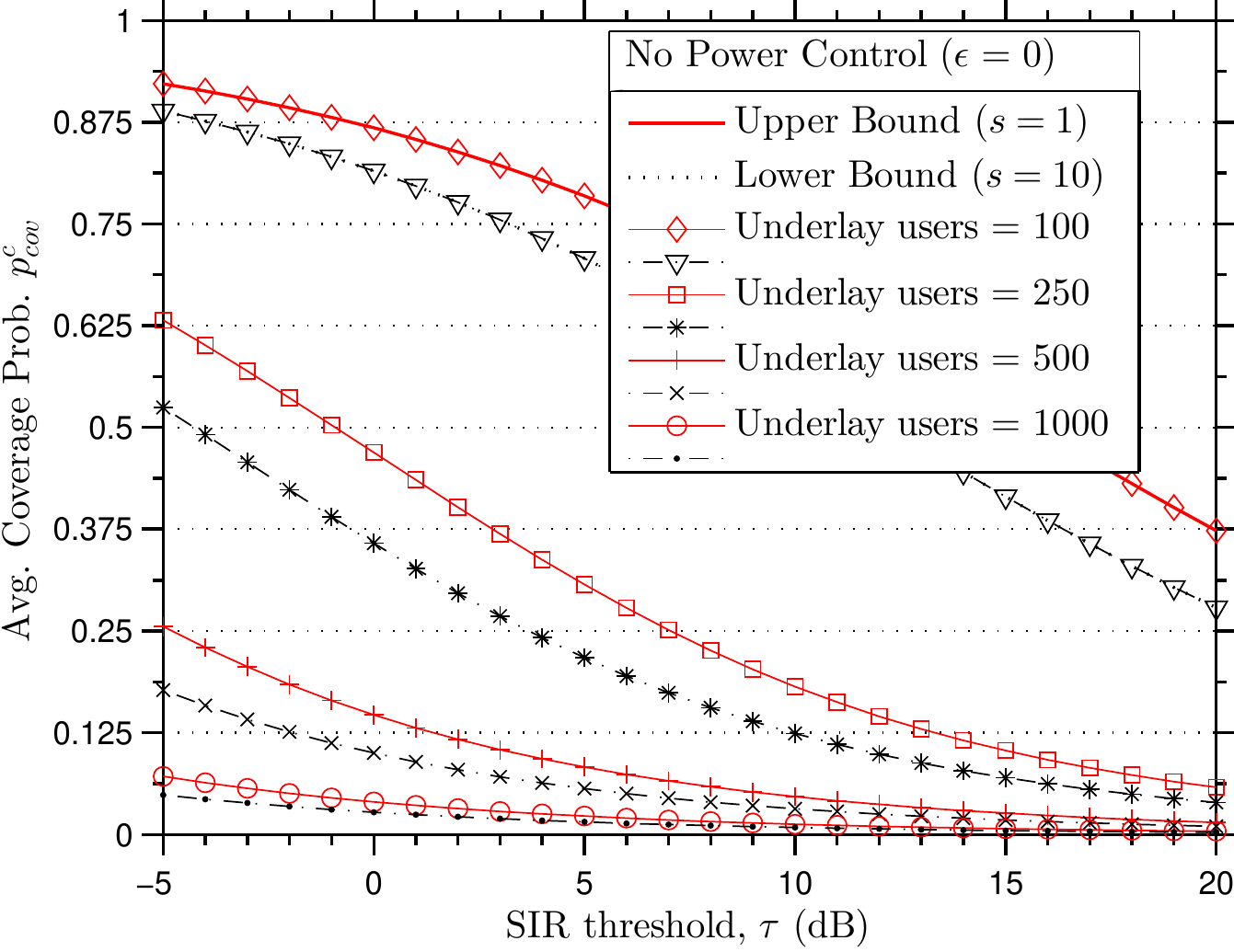}
\caption{Average coverage probability of cellular user for $p_{i}=1$ mW, $p_{c}=0.2$ W, $r_d \leq10$ m, $\alpha=4$, and $\epsilon=0$.}\label{Figure:covsirepsilon0}
\vspace{0mm}
\end{figure}
\begin{figure}[t]
\centering
\includegraphics[width = 1\columnwidth, height = 3 in]{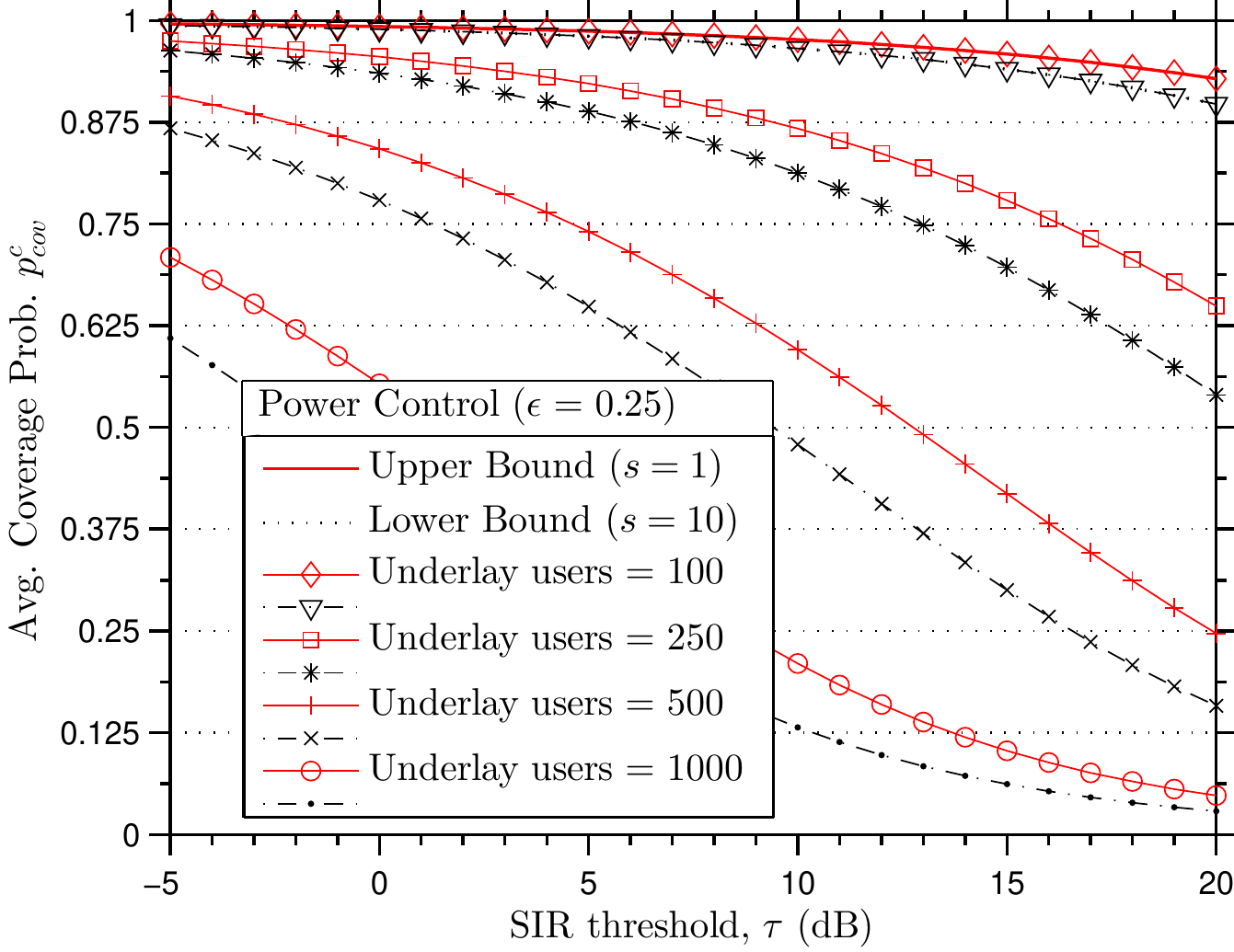}
\caption{Average coverage probability of cellular user for $p_{i}=1$ mW, $p_{c}=0.2$ W, $r_d \leq10$ m, $\alpha=4$, and $\epsilon=0.25$.}\label{Figure:covsirepsilon0p25}
\vspace{0mm}
\end{figure}

In Fig. \ref{Figure:covsirepsilon0}, we analyze average coverage probability of cellular user for different values of SIR threshold $\tau$. The difference between upper and lower bound, for underlay users ($\geq250$) reduces significantly at higher values of $\tau$. Moreover, both bounds converge to a small range of coverage probability, $[0, 0.0625]$. The reverse effect can be seen in case of underlay users ($\leq100$). The bounds converge very slow (at high values of $\tau$) due to reduced number of D2D pairs and subsequently reduced interference. In Fig. \ref{Figure:covsirepsilon0p25}, no convergence to small range of coverage probability can be seen at higher values of $\tau$. This phenomenon follows the intuition. The former case, without any power control, results into interference-dominated environment whereas in later case, simple power control has suppressed accumulated interference of underlay users. The power control effect on coverage probability can be seen by comparing Fig. \ref{Figure:covsirepsilon0} and Fig. \ref{Figure:covsirepsilon0p25}. By incorporating simple power control, dense underlay D2D network can be realized. For example, by setting $\epsilon=0.25$, the lower and upper bounds on coverage probability of cellular user at $\tau=-5$ dB have been improved from $[0.04, 0.07]$ to $[60.7, 70.8]$. For this value of $\epsilon$, the coverage probability of cellular user is ($\geq$) 50\% for underlay users ($\leq$) 250. This shows huge gain in capacity even if half of them are able to share contents between each other while meeting certain SIR threshold\footnote{We assumed maximum of 1000 underlay D2D users. This figure is quite pessimistic as compared to scenarios where 1 million connections$/$Km$^{2}$ is expected in future cellular networks \cite{IMT_2020}.}. 

\begin{figure}[t]
\centering
\includegraphics[width = 1\columnwidth, height = 3 in]{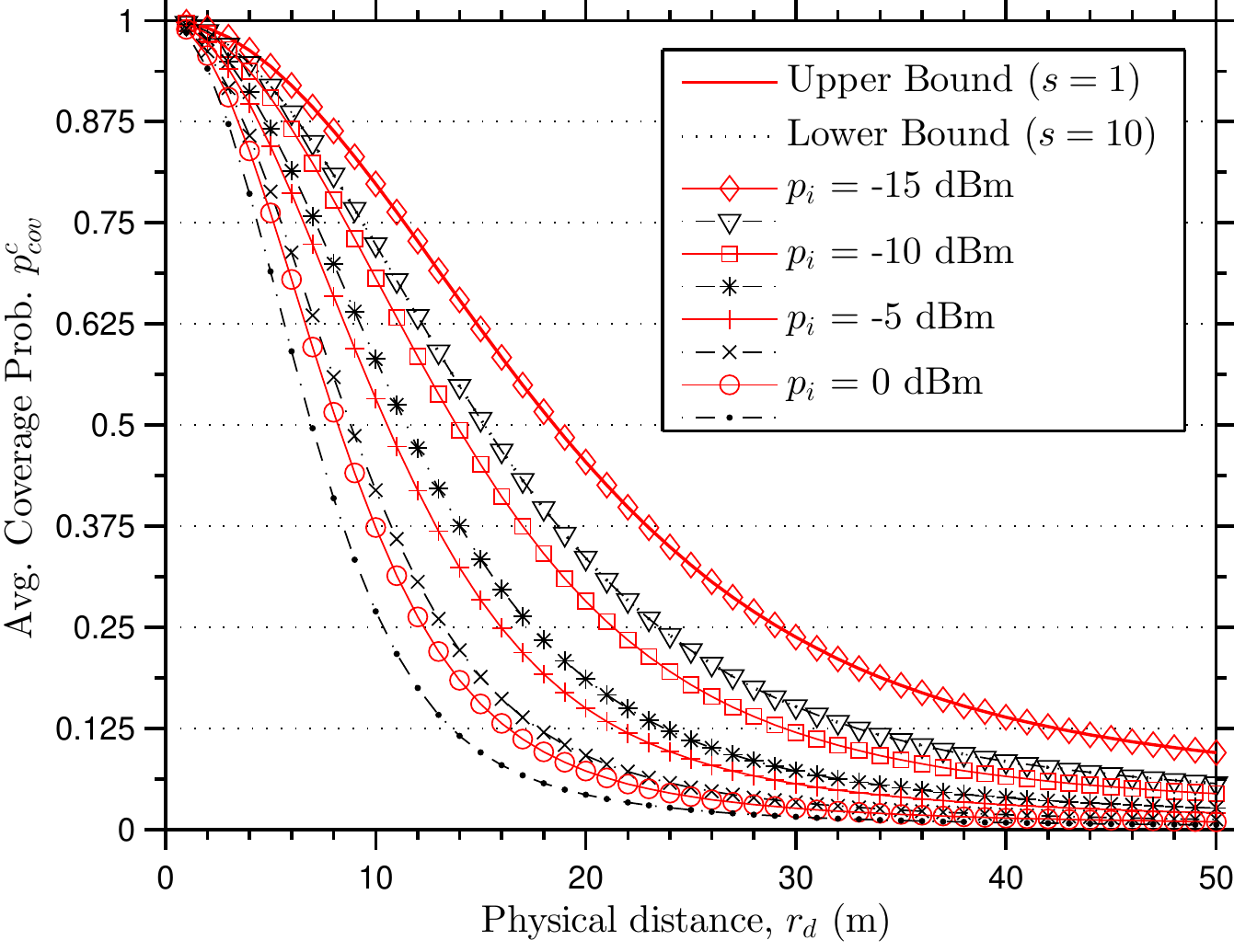}
\caption{Average coverage probability of cellular user for $p_{c}=0.2$ W, $\tau=15$ dB, $\alpha=4$, $\epsilon=0.25$ and 100 underlay D2D users.}\label{Figure:covrd}
\vspace{0mm}
\end{figure}

In Fig. \ref{Figure:covrd}, we analyze the impact of physical distance between D2D users on coverage probability of cellular user. The shortest distance $r_d$ results into reduced path-loss and smaller transmit power. The transmit power of each D2D pair can be different depending upon $r_d$ and required SIR. Therefore, ideally, the transmit power of D2D pairs should be controlled by SBS, however, it requires channel state information of direct link and incurs a lot of signaling overhead. To simplify the analysis, we assume same transmit power of all D2D pairs. Because of the nature of proximity based D2D communication, shorter distances and hence low transmit power is always feasible to contribute minimum interference to the cellular user. We present the impact of $r_d$ for four values of transmit power $p_i=[-15 -10 -5 0]$ dBm. At shorter distances, the direct link establishes line of sight (LOS) communication. The D2D link has an analogy to satellite link however, the distance of former (order of few meters) is negligible as compared to later (order of 36,000 Km). The transmit power of iDirect remote terminal to satellite space segment ranges between -35 and 0 dBm \cite{_installation_2014}. As compared to this distance range, At a distance of few meter, the range of $p_i$ is still very high\footnote{In IP-based satellite communications, iDirect\textsuperscript{\textregistered} is a global leader. The transmit power of iDirect remote terminal to satellite space segment ranges between -35 and 0 dBm \cite{_installation_2014}.}. These values show high interference and hence significantly drop coverage probability of cellular user. For example, increase in $p_i$ from -15 dBm to 0 dBm reduces coverage probability from 79.7\% to 37.3\%. If we reduce transmit power ($\leq-15$ dBm), we can ensure coverage probability of cellular user ($\geq$ 80\%).

\begin{figure}[!htbp]
\centering
\includegraphics[width = 1\columnwidth, height = 3 in]{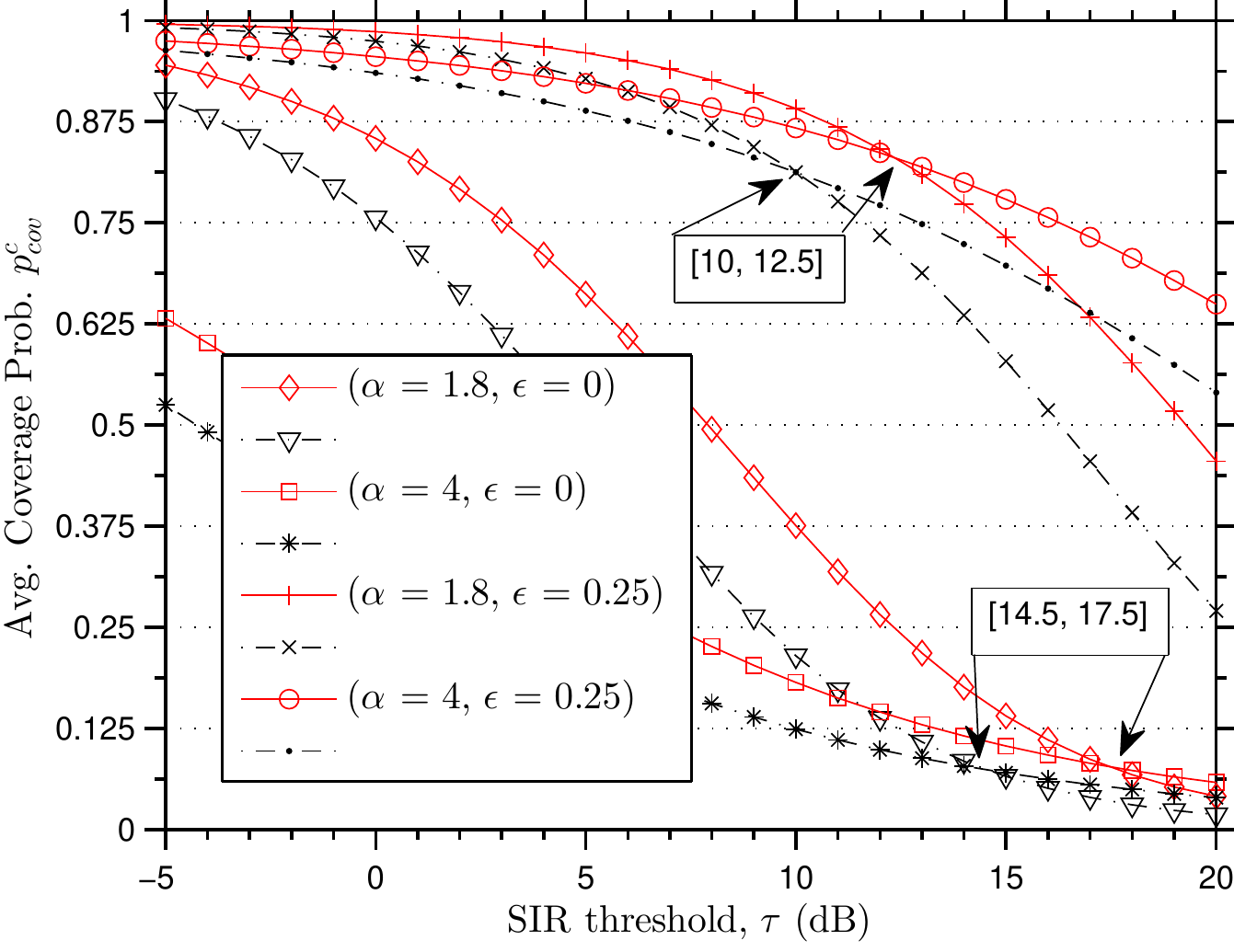}
\caption{Average coverage probability of cellular user for $p_{c}=0.25$ W, $p_{i}=1$ mW, $\tau=15$ dB, $\lambda=0.003$ (250 users), and different ($\alpha$, $\epsilon$).}\label{Figure:covalpha}
\vspace{0mm}
\end{figure}

In Fig. \ref{Figure:covalpha}, we consider two scenarios, indoor ($\alpha=1.8$) and suburban ($\alpha=4$), for two cases of power control ($\epsilon=[0, 0.25]$). In case of no power control, the bounds, at lower $\tau$, are widely spread whereas for higher values, they show cross-over point \cite[Fig. 9, 10]{6516885}. This cross-over point has been shifted to lower values of $\tau$ in case of power control. For example, the cross-over point in these scenarios, in case of upper bounds, has been shifted from ($17.5 \rightarrow 12.5$) by employing power control of 0.25. The similar amount of shift can be seen in case of lower bounds. This shows that in indoor scenarios, the power control provides coverage gain but it rapidly drops as compared to suburban scenario. This validates the intuition since the numerator in (\ref{SIR_AppA}), $I_{A}$, is scaled by big denominator (smaller $\alpha$, higher coverage drop) as compared to very big denominator (higher $\alpha$, smaller coverage drop).

\begin{figure}[!htbp]
\centering
\includegraphics[width = 1\columnwidth, height = 3 in]{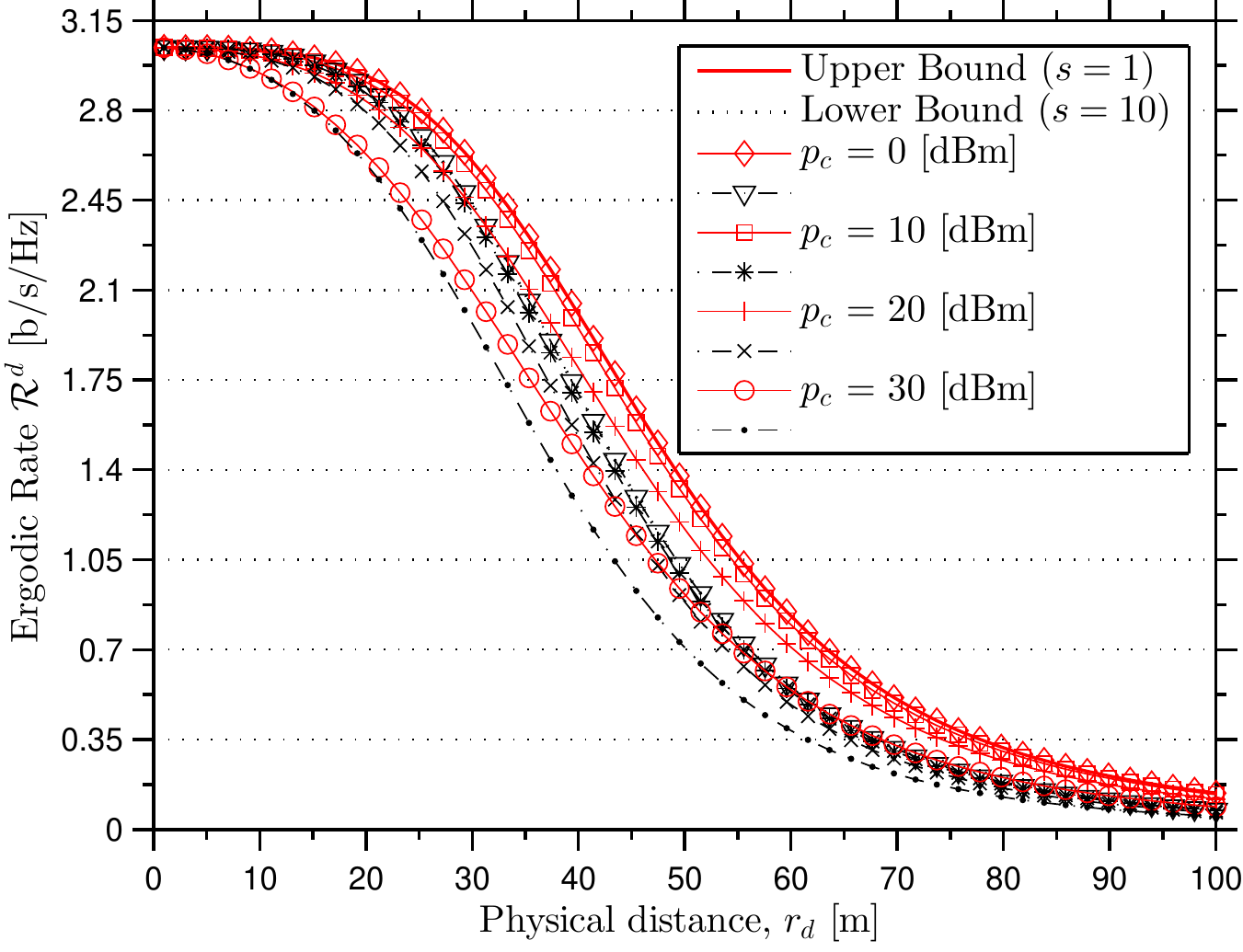}
\caption{Rate of D2D pair for $p_{i}=1$ mW, $\alpha=4$, and variable $r_d$.}\label{Figure:rated2d}
\vspace{0mm}
\end{figure}

The rate of a randomly chosen D2D pair \big(Equ. (\ref{ergrate1})\big) for different values of $r_d$ is shown in Fig. \ref{Figure:rated2d}. Here we can see that for smaller D2D pairing distances, the effect of increased transmission power of cellular user is insignificant. For example, at a distance of $r_d \le 10$, the rate drops very little (i.e., from 3.03 to 2.91 [b/s/Hz]). However, for higher distances ($\ge$ 100), the D2D pair is in the blockage zone.
\section{Conclusions}
In this paper, we characterize interference of underlay D2D network, by considering spatial and social relationship, on average coverage probability of distance-proportional power-controlled cellular user. The spatial and social relations, between D2D nodes, are based on physical distance and Zipf distributed common popular file requests, respectively. The resulting analytic expressions, based on joint spatial and social constraints, have no closed-form expression, however, the lower and upper bounds of common popular file requests have nice convergence in the form of digamma and polygamma functions. We introduce this function as independent marks to each D2D node and apply distance based thinning. Effectively, we apply thinning (physical distance based retention) on Zipf based marked PPP to realize thinned IMPPP process.

The numerical evaluations present the effect of user density, SIR threshold, power control, and physical distance on average coverage probability of cellular user. The analysis show that the user density of potential D2D nodes can be increased by controlling power-control factor of cellular user. This means ultra dense cellular networks can sustain huge capacity demands by up-link power control on cellular user. The other factor is the target SIR threshold of cellular user which can also be ensured by power control. By controlling the transmit power of successful D2D pairs and up-link power control on cellular user, the physical distance for D2D pairing can also be increased to cope up capacity demands.
\balance
\bibliographystyle{IEEEtran}
\bibliography{IEEEabrv,Coverage_Gain_D2D_impp}
\newpage
\begin{IEEEbiography}[{\includegraphics[width=1in,height=1.25in,clip]{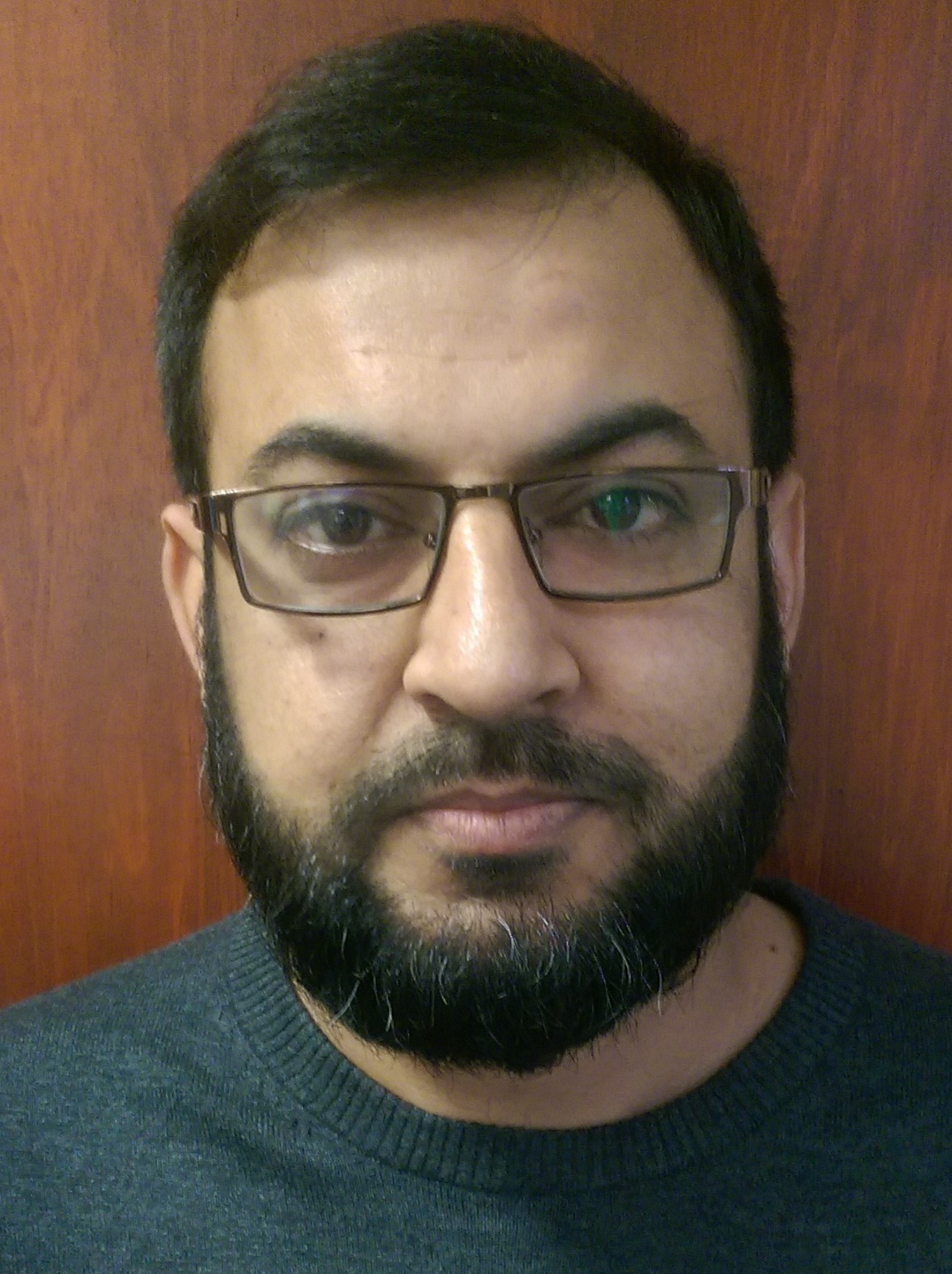}}]
{Hafiz Atta Ul Mustafa} 
received M.Sc. degree in Telecommunication Engineering from National University of Computer \& Emerging Sciences, Pakistan and Ph.D. degree from University of Surrey, UK in 2010 and 2017, respectively. He started as Design Engineer in NexTek Service where he was involved in embedded system design, telecommunication system engineering, product evaluation and project management. Along with that he provided consultancy services in satellite communication systems to Ministry of Defense, Pakistan from 2005 to 2013 (8 years). He is currently working as RF design engineer in Cobbethill Earth Station Ltd. His main research interests include modeling next generation ultra-dense cellular networks, stochastic geometry, point process, RF system design and optimization of communication systems.
\end{IEEEbiography}
\vspace*{-6\baselineskip}
\begin{IEEEbiography}[{\includegraphics[width=1in,height=1.25in,clip]{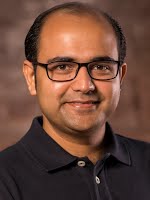}}]
{Muhammad Zeeshan Shakir} 
is an Assistant Professor at the University of the West of Scotland (UWS), UK. Before joining UWS in Fall 2016, he has been working at Carleton University, Canada, Texas A\&M University, Qatar and KAUST, Saudi Arabia on various national and international collaborative projects. Most of his research has been supported by industry partners such as Huawei, TELUS and sponsored by local funding agencies such as Natural Sciences and Engineering Research Council of Canada (NSERC), Qatar National Research Fund (QNRF) and KAUST Global Research Fund (GCR). His research interests include design, development and deployment of diverse wireless communication systems, including hyper-dense heterogeneous small cell networks, Green networks and 5G technologies such as D2D communications, Networked-flying platforms (NFPs) and IoT. He has published more than 75 technical journal and conference papers and has contributed to 7 books, all in reputable venues. He is an editor of 2 research monographs and an author of a book entitled Green Heterogeneous Wireless Networks published jointly by Wiley and IEEE Press. He has been/is serving as a Chair/Co-Chair/Member of several workshops/special sessions and technical program committee of different IEEE flagship conferences, including Globecom, ICC, VTC and WCNC. He is an Associate Technical Editor of IEEE Communications Magazine and has served as a lead Guest Editor/Guest Editor for IEEE Communications Magazine, IEEE Wireless Communications and IEEE Access. He is serving as a Chair of IEEE ComSoc emerging technical committee on backhaul/fronthaul networking and communications. He is a Senior Member of IEEE, an active member of IEEE ComSoc and IEEE Standard Association.
\end{IEEEbiography}
\begin{IEEEbiography}[{\includegraphics[width=1in,height=1.25in,clip]{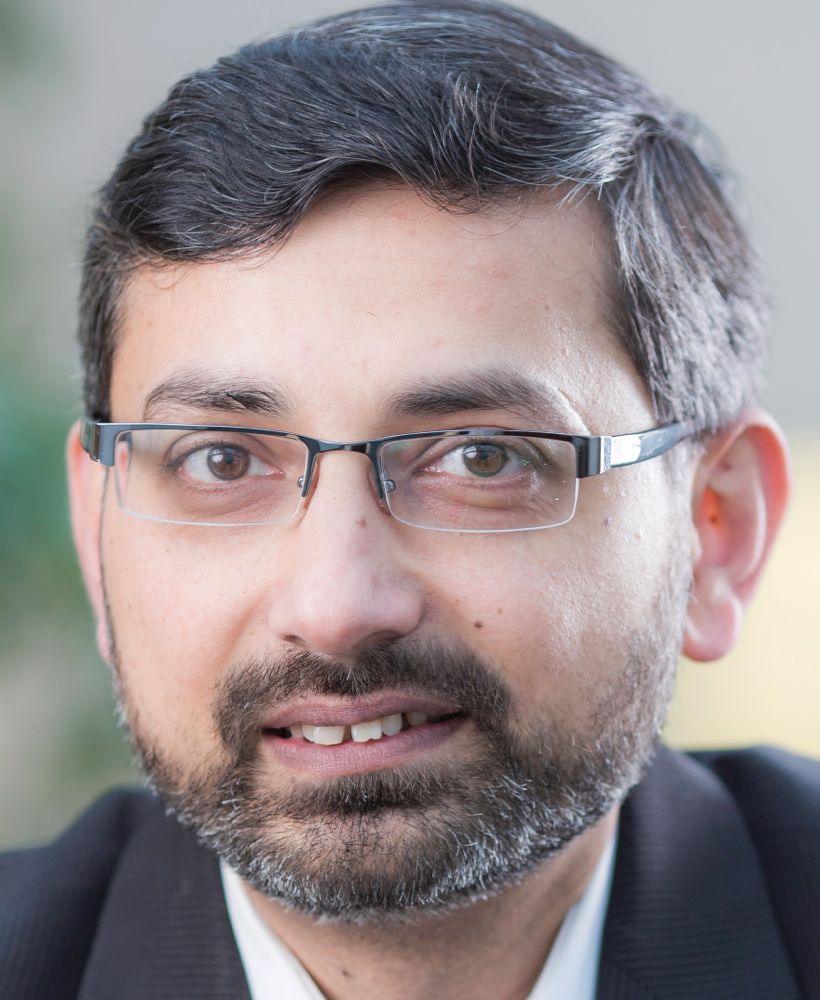}}]
{Muhammad Ali Imran} 
Professor Muhammad Ali Imran (M'03, SM'12) received his M.Sc. (Distinction) and Ph.D. degrees from Imperial College London, UK, in 2002 and 2007, respectively. He is a Professor in Communication Systems in University of Glasgow, Vice Dean of Glasgow College UESTC and Program Director of Electrical and Electronics with Communications. He is an Affiliate Professor at the University of Oklahoma, USA and a visiting Professor at 5G Innovation centre, University of Surrey, UK, where he has worked previously from June 2007 to Aug 2016.
He has led a number of multimillion-funded international research projects encompassing the areas of energy efficiency, fundamental performance limits, sensor networks and self-organising cellular networks. In addition to significant funding from EPSRC, RCUK, Qatar NRF, EU FP7/H2020, he has received direct industrial funding from leading industries in Communications: Huawei, Sony, IBM, DSTL, British Telecom, He also lead the new physical layer work area for 5G innovation centre at Surrey. He has a global collaborative research network spanning both academia and key industrial players in the field of wireless communications. He has supervised 25+ successful PhD graduates and published over 300 peer-reviewed research papers.  He secured first rank in his B.Sc. and a distinction in his M.Sc. degree along with an award of excellence in recognition of his academic achievements conferred by the President of Pakistan. He has been awarded IEEE Comsoc’s Fred Ellersick award 2014, Sentinel of Science Award 2016, FEPS Learning and Teaching award 2014 and twice nominated for Tony Jean’s Inspirational Teaching award. He is a shortlisted finalist for The Wharton-QS Stars Awards 2014, Reimagine Education Awards 2016 for innovative teaching and VC’s learning and teaching award in University of Surrey. He is a senior member of IEEE and a Senior Fellow of Higher Education Academy (SFHEA), UK.
He has given an invited TEDx talk (2015) and more than 10 plenary talks, several tutorials and seminars in international conferences and other institutions. He has taught on international short courses in USA and China. He is the co-founder of IEEE Workshop BackNets 2015 and chaired several tracks/workshops of international conferences. He is an associate Editor for IEEE Communications Letters, IEEE Open Access and IET Communications Journals.
\end{IEEEbiography}
\vspace*{-2\baselineskip}
\begin{IEEEbiography}[{\includegraphics[width=1in,height=1.25in,clip]{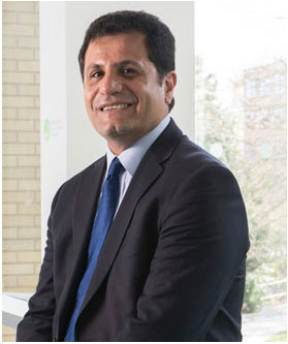}}]
{Rahim Tafazolli} 
(SM'09) is a Professor of Mobile and Satellite Communications, Director of Institute for Communication Systems (ICS), and the founder and Director of 5G Innovation Centre (5GIC) at the University of Surrey, UK.
He has over 25 years of experience in digital communications research and teaching.  He has authored and co-authored more than 500 research publications. He is regularly invited to deliver keynote talks and distinguished lectures to International conferences and workshops. He is a co-inventor of more than 30 granted patents, all in the field of digital communications.  He is regularly invited by many governments for advise on 5G technologies.
He is a Fellow of Wireless World Research Forum (WWRF) in recognition of his personal contributions to the wireless world as well as heading one of Europe’s leading research groups.
\end{IEEEbiography}
\end{document}